\documentclass[aps,prb,twocolumn,amsmath,amssymb,groupedaddress,floatfix,longbibliography]{revtex4-2}
\setcitestyle{square,numbers}

\usepackage{graphicx}
\usepackage{bm}
\usepackage{color}
\usepackage{epstopdf}
\usepackage{comment}
\usepackage[plainpages=false,pdfpagelabels,colorlinks=true,linkcolor=red,urlcolor=blue,citecolor=blue,pdftitle={Title},pdfauthor={},pdfdisplaydoctitle=true,pdfduplex=DuplexFlipLongEdge]{hyperref}

\begin{document}

\title{Pairing and charge distribution in Emery ladders \\ preserving the ratio of Cu to O atoms}
\author{G\"{o}kmen Polat}
\author{Eric Jeckelmann}
\affiliation{Leibniz Universit\"{a}t Hannover, Institute of Theoretical Physics,  Appelstr.~2, 30167 Hanover, Germany}

\date{\today}

\begin{abstract}
We study the Emery model (three-band Hubbard model) for superconducting cuprates on three distinct ladder-like lattices that are supercells of the CuO$_2$ plane 
and thus preserve the ratio of copper to oxygen atoms.
Using the density-matrix renormalization group method we confirm that these Emery ladders are charge-transfer insulators for the hole concentration corresponding to undoped cuprates 
but become Luther-Emery liquids with enhanced pairing correlations upon doping. 
The preservation of the Cu to O ratio allows us to study the distribution of charges between these atoms in the Luther-Emery phase. We show that these Emery ladders
can describe the relations between charge distribution, pairing strength, and interactions that have been observed in the Emery model
on two-dimensional clusters and in experiments.
\end{abstract}

\maketitle

\section{Introduction} 
Understanding cuprate high-temperature superconductors remains a theoretical challenge forty years after their discovery~\cite{Bednorz1986}. 
The  two-dimensional (2D) Emery model (or three-band Hubbard model)~\cite{Emery1987a} is generally believed to describe the low-energy electronic properties 
of these materials~\cite{Dagotto1994}. As this model includes both copper and oxygen orbitals of the CuO$_2$ planes, it
allows one to describe undoped cuprates as charge-transfer insulators  and to investigate the charge distribution between copper and oxygen atoms in doped cuprates~\cite{Jeckelmann1998c,Kowalski2021,St-Cyr2025,Reaney2025,Peng2025}.
Given the difficulty of studying  2D strongly correlated systems, quasi-one-dimensional (1D) subsystems called ladders have often been investigated
with the aim of shedding light on the properties of their 2D parents~\cite{Dagotto1992,Dagotto1996}. 
The advantage of ladder systems is  the availability of well-established methods for 1D correlated models such as the density-matrix renormalization group (DMRG)~\cite{White1992b,White1993a,Schollwoeck2005,Jeckelmann2008a,Schollwoeck2011}.
The Emery model has been studied on two-leg ladders using the DMRG method for nearly 30 years~\cite{Jeckelmann1998c,Nishimoto2002b,Nishimoto2009}.
Recently, this approach has been used~\cite{Song2021,Song2023,Jiang2023a,Jiang2023b,Yang2024,Polat2025}
to investigate the existence of a Luther-Emery phase with enhanced pairing correlations~\cite{Giamarchi2003}, which is the 1D counterpart of the superconducting phase in higher dimensions.

In our previous work~\cite{Polat2025}, we demonstrated the existence of a robust Luther-Emery phase with enhanced pairing correlations in the three common
ladder structures used in 
other studies~\cite{Jeckelmann1998c,Nishimoto2002b,Nishimoto2009,Song2021,Song2023,Jiang2023a,Jiang2023b,Yang2024}, which we called 3-chain ladder, 5-chain ladder, and 4-chain tube.
However, the ratio of Cu to O orbitals is 2:3 and 2:5 in the 3-chain and 5-chain ladders, respectively, which differs 
significantly from the ratio 1:2 of the 2D CuO$_2$ lattice. The 4-chain tube has the right ratio but pairing occurs only
if the hopping terms between Cu and O orbitals are anisotropic. Thus it is questionable whether these common ladders can approximate
the essential properties of  the isotropic 2D CuO$_2$ lattice, especially,  the charge distribution between Cu and O atoms.
  
 In this work we propose three two-leg ladder structures with a Cu to O ratio of 1:2. Each ladder can be seen as a supercell of the 2D CuO$_2$ lattice. 
A drawback is that they have lower symmetries than the common ladder structures. 
We study the Emery model on the proposed  ladder structures and show that they exhibit the low-energy properties expected for cuprates with realistic parameters.  
In particular, they are charge-transfer insulators for the hole concentration corresponding to undoped cuprates but become Luther-Emery liquids upon doping. 
This allows us to study  the distribution of charges between copper and oxygen atoms as well as the pairing strength
as a function of doping and Hamiltonian parameters.
We find systematic relations in agreement with experimental results~\cite{Rybicki2016,Jurkutat2023} and theoretical predictions for 2D clusters~\cite{Kowalski2021,St-Cyr2025,Reaney2025,Peng2025}. 

The structure of the paper is as follows: In Sec.~\ref{sec:model}, we introduce the ladder structures, the Emery model and the method used.
We discuss the properties of the ladders at hole concentration corresponding to undoped cuprates in Sec.~\ref{sec:undoped}.
In Sec.~\ref{sec:doped} we investigate their pairing properties upon electron and hole doping.
The charge distribution on the oxygen and  copper orbitals is discussed in Sec.~\ref{sec:charge}.
Finally, we present our conclusion in Sec.~\ref{sec:conclusion}.

\section{Model and method \label{sec:model}} 

\subsection{Ladder structures}

The low-energy electronic excitations of superconducting cuprates are confined in their CuO$_2$ planes~\cite{Dagotto1994,Scalapino2012}.
Thus much effort has been devoted to understanding the properties of strongly correlated electron systems on 2D lattices.
However, most numerical approaches can only handle small subsystems of 2D lattices accurately.
In particular,  the DMRG method has been very successfully applied to ladder lattices.
The original Hubbard ladders~\cite{Noack1994} are simply made of two coupled 1D Hubbard chains and they can be seen as narrow strips cut out of a square lattice.

\begin{figure}
\includegraphics[width=0.35\textwidth]{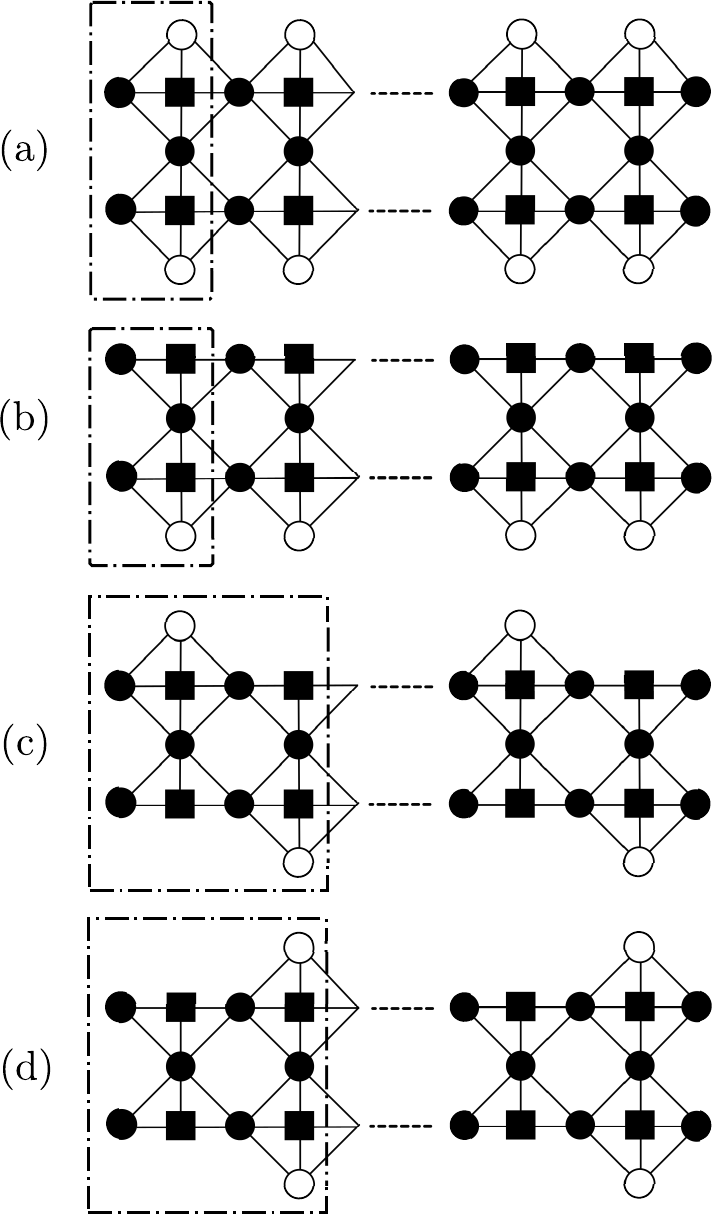}
\caption{\label{fig01} Schematic of various two-leg ladder lattice structures. Squares and circles represent copper $d$ orbitals and oxygen $p$ orbitals, respectively. Solid black circles and squares are included 
in all lattice structures. White circles denote the outer oxygen $p$ orbitals that are included in specific ladders only. Dashed-dotted lines indicate the unit cells. Solid lines represent the hopping terms between orbitals in the Emery model~(\ref{eq:hamiltonian}).
(a) Three-chain ladder made of the solid black symbols, five-chain ladder made of all symbols, and four-chain tube, where the upper and lower white  circles represent the same oxygen atoms,
(b) t-ladder with translation symmetry, (c) g-ladder with glide symmetry, and (d) r-ladder with reflection symmetry.
}
\end{figure}

For the 2D CuO$_2$ lattice, however, there are many possibilities of cutting out a strip of Cu and O atoms.
Several ladder structures for cuprates are shown in Fig.~\ref{fig01}.
 All examples  in this figure are two-leg ladders composed of two chains of copper atoms and a variable number of oxygen atoms. 
 Figure~\ref{fig01}(a) illustrates three common ladder structures that have already been studied in the framework of the Emery model~\cite{Jeckelmann1998c,Nishimoto2002b,Nishimoto2009,Song2021,Song2023,Jiang2023a,Jiang2023b,Yang2024,Polat2025}:
 the three-chain ladder,  the five-chain ladder, and the four-chain tube.
Note that they possess a reflection symmetry in the rung direction (vertical or $y$-direction) and that infinitely long ladders also have a discrete translational invariance in the leg direction (horizontal or $x$-direction).

 Although the study of these  ladders has  yielded much information about the properties of the Emery model,
 their adequacy for investigating the properties of 2D systems is questionable~\cite{Polat2025}.
 Indeed, these ladders cannot be seen as a supercell of the 2D CuO$_2$ lattice, i.e we do not recover
 the correct 2D lattice structure by repeating periodically one of these  ladder structures.
 An obvious inconsistency is that the ratio of Cu to O atoms is 2:3 and 2:5 in the 3-chain and 5-chain ladders, respectively, instead of the ratio 1:2 in the 2D CuO$_2$ lattice.
 This is an important issue when studying the charge redistribution between Cu and O atoms as discussed in Sec.~\ref{sec:charge}.
 The four-chain tube has the correct ratio of both atoms but as a result of the periodic hopping terms in the rung direction, it exhibits a Luther-Emery phase only for hopping terms that are stronger in 
 the leg direction than in the rung direction~\cite{Polat2025}, which is not compatible with the symmetry of the 2D CuO$_2$  lattice under 90 degree rotations.
 
However,  various two-leg ladder structures that are supercells and thus have the correct ratio of Cu to O atoms can be cut out of the 2D CuO$_2$ lattice.
Three examples are shown in Fig.~\ref{fig01}(b) to (d). Obviously, they possess a lower symmetry than the common ladders shown in Fig.~\ref{fig01}(a).
The t-ladder in Fig.~\ref{fig01}(b) preserves the discrete translational invariance but breaks the reflection symmetry. The g-ladder in Fig.~\ref{fig01}(c) possesses 
a glide symmetry instead of the separate translation and reflection symmetries. Thus its unit cell is twice as large as for the common ladder structures.
 Finally, the r-ladder in Fig.~\ref{fig01}(d) preserves the reflection symmetry in the rung direction but its periodicity in the leg direction is also twice as 
 long as for the common ladder structures. The symmetries and unit cells are summarized in Table~\ref{table1}.

 \begin{table}
\centering
\begin{tabular}{|c |c |c |c |c |c |c |c|} 
 \hline
 Ladder  & Unit cell & \hspace{1mm}  $T_{1}$ \hspace{1mm}  & \hspace{1mm}  $T_{2}$ \hspace{1mm}  &  \hspace{1mm}  $R$  \hspace{1mm}  \\ [0.5ex] 
 \hline\hline
  	3-chain		& Cu$_2$O$_3$ 	& \checkmark 	&  \checkmark 	& \checkmark	\\
   	5-chain 		& Cu$_2$O$_5$	& \checkmark	& \checkmark 	& \checkmark 		\\ 
	tube         & Cu$_2$O$_4$  & \checkmark	& \checkmark	& \checkmark 		\\ 
	t		& Cu$_2$O$_4$ 	& \checkmark	& \checkmark  	& - 		  \\ 
        g		& Cu$_4$O$_8$ 	&  -	& \checkmark  	& -  		 \\ 
	r		& Cu$_4$O$_8$	&  -	& \checkmark  	& \checkmark   \\
 \hline
\end{tabular}
\caption{Unit cell and symmetries of the six ladder structures illustrated in Fig.~\ref{fig01}. $T_1$ indicates that an infinitely long ladder is
invariant under discrete translations in the leg direction with a period equal to the lattice constant of the 2D CuO$_2$ lattice. 
$T_2$ corresponds to  a twice longer lattice period.
$R$ indicates a  reflection symmetry in the rung direction. 
\label{table1}
}
\end{table}

\subsection{Hamiltonian \label{sec:hamiltonian}}

The Emery model~\cite{Emery1987a,Dagotto1994} extends the Hubbard model to describe holes in the pertinent orbitals of a 2D CuO$_2$ lattice.
The model incorporates one $d_{x^2-y^2}$ orbital for each copper atom and one $p_x$  or $p_y$ orbital for each oxygen atom.
We apply this model to the  two-leg ladders shown  in Fig.~\ref{fig01}. 
The system size $L$ is defined as the number of $d$ orbitals on a single leg. This is also  the number of rungs in the ladder. 
The undoped state corresponds to a filling of $N=2L$ holes. The doping concentration is defined as $\delta = N/(2L)-1$.
Thus $\delta=0$ corresponds to an undoped ladder, while $\delta>0$ when the system is hole-doped but holes are removed ($\delta<0$) upon electron doping.

The Hamiltonian is 
\begin{eqnarray}\label{eq:hamiltonian}
H=&-& t_{dpx} \sum_{\left\langle ij  \right\rangle, \sigma } \left( p^{\dagger}_{xi\sigma} d^{\phantom{\dagger}}_{j\sigma} + \text{H.c.}\right)   \nonumber \\
&-& t_{dpy} \sum_{\left\langle ij  \right\rangle, \sigma } \left( p^{\dagger}_{yi\sigma} d^{\phantom{\dagger}}_{j\sigma} + \text{H.c.}\right) \nonumber \\
&-& t_{pp} \sum_{\left\langle ij \right\rangle, \sigma} \left( p^{\dagger}_{xi\sigma} p^{\phantom{\dagger}}_{yj\sigma} + \text{H.c.} \right)  \\
&+& \varepsilon \sum_{\alpha,i,\sigma} n^p_{\alpha i \sigma} \nonumber \\
&+& U_d \sum_i n^d_{i \uparrow} n^d_{i \downarrow} + U_p \sum_{\alpha,i} n^p_{\alpha i \uparrow} n^p_{\alpha i \downarrow} .  \nonumber 
\end{eqnarray}
where $c^{\phantom{\dagger}}_{i\sigma}$ is the annihilation operator and $c^{\dagger}_{i\sigma}$ is the creation operator for a hole with spin $\sigma$ in a $d$ orbital ($c=d$), 
a $p_x$ orbital ($c=p_x$), or a $p_y$ orbital ($c=p_y$).
The hole number operators are defined by $n^d_{i\sigma} = d_{i\sigma}^{\dagger} d_{i\sigma}^{\phantom{\dagger}}$ and $n^p_{\alpha i\sigma} = p_{\alpha i\sigma}^{\dagger} p_{\alpha i\sigma}^{\phantom{\dagger}}$
with $\alpha=x,y$.
The first three terms in the Hamiltonian represent hopping processes between neighbor orbital pairs.
The first sum  runs over all pairs $\langle ij  \rangle$ made of nearest-neighbor $d$-$p_x$ orbitals (represented by horizontal lines in Fig.~\ref{fig01}) while 
the second sum is  over all nearest-neighbor $d$-$p_y$ pairs (vertical lines).
The third  sum runs over all nearest-neighbor  $p_x$-$p_y$ orbital pairs (represented by diagonal lines in Fig.~\ref{fig01}).
The fourth term in the Hamiltonian~(\ref{eq:hamiltonian})  is the energy difference between a single hole in a $p$ orbital and one in a $d$ orbital.
The sum over the indices $i$ and $\alpha$ indicates a sum over all $p$ orbitals.
The last two terms represent the Coulomb repulsion between two holes in the same $d$ or $p$ orbital, respectively,
and their sums run over all corresponding orbitals.

The Hamiltonian~(\ref{eq:hamiltonian})  comprises  six parameters. First, there are the energy difference $\varepsilon > 0 $ as well as the on-site Coulomb repulsion strengths $U_d > 0$ and $U_p \geq 0$. 
Second, there are the interatomic hoppings $t_{dpx} > 0$  along the chains between $d$ and $p_x$ orbitals, $t_{dpy} > 0$  along the rungs between $d$ and $p_y$ orbitals, 
and $t_{pp}$ between $p_x$-$p_y$ orbital pairs. 
The orbital phases are chosen such that all hopping terms have a constant sign.
We use the energy unit $t_{dpx}=1$ for numerical results throughout.
For superconducting cuprates $t_{dpx}=t_{dpy}$ is of the order of 1 eV.

Numerous different  values can be found in the literature for these Hamiltonian parameters~\cite{Emery1987a,Ogata1988,Dagotto1994,Martin1996,Sheshadri2023,Jeckelmann1998c,Nishimoto2002b,Nishimoto2009,Song2021,Song2023,Jiang2023a,Jiang2023b,Yang2024,Polat2025}.
Here we are mostly interested in the pairing properties of the Emery ladders presented in Fig.~\ref{fig01}
and the pair binding energy (PBE) is the most important observable in our investigation.
 Therefore, for each ladder structure we have selected a parameter set that yields a significant PBE for two doped holes.
The PBE is defined as
\begin{eqnarray}\label{eq:Epb}
    E_{pb} &=& 2E_0\left( N_{\uparrow} \pm 1,N_{\downarrow} \right) - E_0\left( N_{\uparrow} \pm 1,N_{\downarrow} \pm 1 \right) \nonumber \\
    &-& E_0 \left( N_{\uparrow},N_{\downarrow} \right) 
\end{eqnarray}
where $E_{0}\left( N_{\uparrow}, N_{\downarrow} \right)$ is the ground-state energy of the ladder system with $N_{\sigma}$ holes of spin $\sigma$. 
The plus sign is used for hole doping and the minus sign for electron doping.
A positive PBE indicates that doped particles  (holes or electrons) gain energy by building a bound pair.
In this study we will refer to this case as ``pairing''. 

To find parameters for a given ladder, we start from typical values found in the literature~\cite{Dagotto1994}
and perform a blind search maximization of the PBE over the free Hamiltonian parameters
$\varepsilon$, $U_d$, $U_p$, $t_{dpy}$, and $t_{pp}$. 
Only PBE for two doped holes in ladders of length $L=24$ are considered.
Moreover, we require the undoped ladder to be a charge-transfer insulator~\cite{Zaanen1988}
with the charge ratio of around 3:7 on the $p$ and $d$ orbitals (see Sec.~\ref{sec:undoped}).
When the PBE is positive, we  have always found that it is maximal for
$U_p=0$ and $t_{pp}=0$. 
Therefore, we only show results for these values of $U_p$ and $t_{pp}$ in this work.
The other three parameters are slightly different for each ladder structure, confirming the influence
of the oxygen sites. However, to make comparison easier we have selected a single parameter set
for the (t,g,r)-ladders.
The selected parameters are listed in Table~\ref{table2}.

\begin{table}
\centering
\begin{tabular}{ |c |c |c |c |c |c |}
 \hline
  Ladder &  \hspace{2mm}  $\varepsilon$ \hspace{2mm}  & \hspace{0.mm}  $t_{dpy}$  \hspace{0mm}  & \hspace{1.7mm} $U_d$  \hspace{1.3mm} & \hspace{1.mm}  $n_d$ \hspace{1.mm} & \hspace{2.6mm} $E_{pb}$ \hspace{1.0mm} \\ [0.5ex]
 \hline\hline
       t        & 2.00    & 1.1  &  8    & 0.69 & 0.043 \\
    g        & 2.00     & 1.1      & 8 &  0.69 & 0.027 \\
     r        & 2.00    & 1.1      & 8  & 0.69 & 0.040 \\
           \hline
        3-chain        &   2.25      &   1.1  & 4  & 0.72 & 0.057\\
        5-chain         &   2.77    & 1.08  & 4.62  & 0.75 & 0.042\\
      tube            &   2.25 & 0.8     & 4 & 0.72 & 0.058 \\
 \hline
\end{tabular}
\caption{Selected parameters $\varepsilon$, $t_{dpy}$, and  $U_d$ for the Hamiltonian~(\ref{eq:hamiltonian}) on each ladder structure.
These parameters are used for most results presented here. $n_d$ is the probability of finding a hole in the $d$ orbitals in undoped ladders. 
$E_{pb}$ is the pair binding energy~(\ref{eq:Epb}) for two doped holes in ladders of length $L=24$.
\label{table2}}
\end{table} 

In the noninteracting limit ($U_d=U_p=0$) one can diagonalize the Hamiltonian~(\ref{eq:hamiltonian}) exactly for the three common ladder structures
in Fig.~\ref{fig01}(a)~\cite{Polat2025}. We have not been able to carry out this calculation analytically for the (t,g,r)-ladders
 because of their lower symmetries. Nevertheless, one can easily compute their single-particle energy spectrum numerically.
We have checked that two single-particle bands cross the Fermi energy in lightly doped ladders
with the parameters $\varepsilon$ and $t_{dpy}$ listed in Table~\ref{table2}.
According to field theory~\cite{Giamarchi2003} the corresponding interacting system should be a Luther-Emery liquid.
In our previous study of the common Emery ladders~\cite{Polat2025} we found that  this generic field-theoretical prediction
about two-leg ladders was correct. We can confirm here that this prediction is also correct for the (t,g,r)-ladders,
at least for the parameters in Table~\ref{table2}.

It is well known that the 2D Emery model and the 2D Hubbard model on a square lattice describe the same low-energy physics 
for a range of Hamiltonian parameters and band fillings based on the Zhang-Rice mapping ~\cite{Dagotto1994,Zhang1988}.
Similarly, one can map the common Emery ladders in Fig.~\ref{fig01} onto a Hubbard ladder,
which is known to exhibit pair binding and enhanced pairing correlations~\cite{Noack1994,Noack1997,Jeckelmann1998c},
although additional hopping or interaction terms are necessary to recover the low-energy physics of hole-doped Emery ladders~\cite{Jeckelmann1998c,Polat2025}.
The mapping is more complex for the ladder structures in Fig.~\ref{fig01}(b) to (d).
In the case of the t-ladder one obtains a two-leg ladder with different parameters on each leg.
This asymmetric Hubbard ladder has been  investigated recently~\cite{Yoshizumi2009,Abdelwahab2015, Abdelwahab2018b, Abdelwahab2023}.
It also exhibits pair binding and enhanced pairing correlations~\cite{Abdelwahab2023}.
The g-ladder and the r-ladder can be mapped onto generalized Hubbard ladders with parameters that alternate
from one site to another or from one rung to another, respectively. The pairing properties of such models have not been studied
as far as we know.

\subsection{DMRG}
For our investigation of the  Emery model  on two-leg ladders 
we compute the ground-state properties of the Hamiltonian (\ref{eq:hamiltonian}) using the DMRG method~\cite{White1992b,White1993a,Schollwoeck2005,Jeckelmann2008a,Schollwoeck2011}.
This method has been successfully applied to Emery ladders in several studies~\cite{Jeckelmann1998c,Nishimoto2002b,Nishimoto2009,Song2021,Song2023,Jiang2023a,Jiang2023b,Yang2024,Polat2025}.
Our DMRG calculations are carried out using the  C$^{++}$ version of the ITensor Software Library 
for Tensor Network Calculations~\cite{ITensor}.
We have validated our program using  exact diagonalizations of small clusters~\cite{Weisse2008} and an old DMRG code employed in previous studies~\cite{Jeckelmann1998c,Nishimoto2002b}.
We use bond dimensions up to $9000$ resulting in discarded weights  of the order of $10^{-8}$ or smaller. 
We always start the DMRG calculation with a lower bond dimension and add a noise term (typically $10^{-5}$), which is progressively reduced down to zero
as the bond dimension increases~\cite{White2005}.
We investigate systems with even ladder lengths $L$ up to $L=40$.
To obtain a higher DMRG accuracy, open boundary conditions are used  in the leg-direction but
we always include oxygen sites at both ladder ends, as illustrated in Fig.~\ref{fig01},
because we found that this configuration resulted in weaker boundary effects in our previous work~\cite{Polat2025}.
Nevertheless, finite-size effects may be significant and are discussed when we present our results.
In our DMRG calculations the total number of $d$ orbitals is $2L$ while the total number of $p$ orbitals is $4L+2$ for the (t,g,r)-ladders and the 4-chain tube
while it is $3L+2$ and $5L+2$ for the 3-chain and 5-chain ladders, respectively.

\section{Undoped ladders \label{sec:undoped}}

  \begin{figure}
\includegraphics[width=0.48\textwidth]{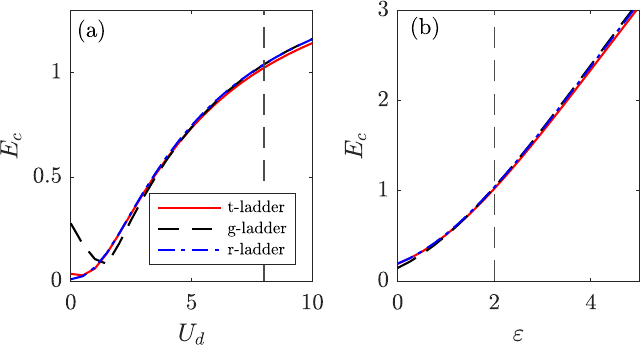}
\caption{\label{fig02} Charge gaps $E_c$ (\ref{eq:charge_gap}) as a function of the Hamiltonian parameters $U_d$ (left panel) and $\varepsilon$ (right panel) for undoped  (t,g,r)-ladders
with length $L=24$.
The other Hamiltonian parameters are given in Table~\ref{table2}. Vertical dashed lines indicate the selected values  $U_d=8$ and $\varepsilon=2$, respectively.}
\end{figure}

The undoped parents of superconducting cuprates are antiferromagnetic charge-transfer insulators~\cite{Dagotto1994,Scalapino2012, Zaanen1988}.
This state is reproduced by the undoped 2D Emery model for a wide range of parameters.
It is also known that the common Emery ladders are charge-transfer insulators
with short-range antiferromagnetic order for appropriate parameters~\cite{Jeckelmann1998c,Nishimoto2002b}.
In this section we demonstrate  that the (t,g,r)-ladder structures have similar properties
for parameters around those listed in Table~\ref{table2}.

 The gap for charge excitation can be calculated from the ground-state energy $E_0$ for different number of particles using
\begin{align}
E_c=& [ E_0(N_\uparrow+1,N_\downarrow+1)  
+E_0(N_\uparrow-1,N_\downarrow-1)]/2 \nonumber \\
&-E_0(N_\uparrow,N_\downarrow) \label{eq:charge_gap} .
\end{align}
In Fig.~\ref{fig02},  we show the charge gap as a function of the Hamiltonian parameters $U_d$ and $\varepsilon$.
Clearly, one can obtain charge gaps of the order of $t_{dpx}=1$ for reasonable parameter values, as found for the common ladders in previous works~\cite{Jeckelmann1998c,Nishimoto2002b}.
Assuming $t_{dpx} \approx$ 1eV these charge gaps are compatible with experimental values for cuprate ladder materials~\cite{Popovi2000,Gozar2001}.
Note that the charge gap of the g-ladder does not vanish for $U_d=0$ in Fig.~\ref{fig02}(a). Thus our DMRG result agrees with the single-particle spectrum
of the Hamiltonian~(\ref{eq:hamiltonian}) in the noninteracting limit, which corresponds to a band insulator for this undoped ladder structure.
By contrast, the single-particle spectrum of the noninteracting  t-ladders and r-ladders is gapless at the Fermi energy corresponding to undoped Emery ladders,
in agreement with the vanishing charge gap calculated with DMRG for $U_d=0$.

The local hole density on orbital $i$  is defined
as the ground-state expectation value of the corresponding hole number operator,
$\rho_i = \sum_{\sigma} \langle n^c_{i\sigma} \rangle$ with $c=d,p$. This density provides us with useful information about the charge distribution in the
ground state and the low-energy charge excitations. 
First, we can define  the average densities $n_d$ 
over all $d$ orbitals and $n_p$ over all $p$ orbitals. In undoped Emery ladders 
$n_d + \gamma  n_p = 1$, where $\gamma$ is the ratio of $p$ to $d$ orbitals in the unit cell.
For the (t,g,r)-ladders and the four-chain tube $\gamma=2$  but $\gamma=3/2$ and $5/2$
for the three-chain and five-chain ladders, respectively.
Experimentally, the charge ratio on the $p$ and $d$ orbitals is around 3 to 7 for YBa$_2$Cu$_3$O$_6$ \cite{Jurkutat2023}.
We have selected Hamiltonian parameters so that $\gamma n_p/n_d$ reproduces this ratio.
The values of $n_d$ are shown in Table~\ref{table2}.

 \begin{figure}
\includegraphics[width=0.48\textwidth]{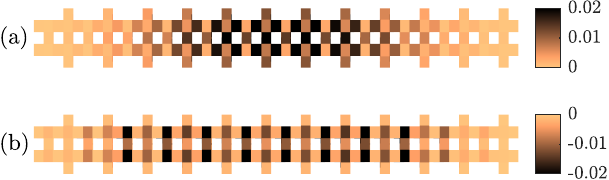}
\caption{\label{fig03} Normalized hole density variation~(\ref{eq:local_density}) of the r-ladder with length $L=24$ and the parameter set in Table~\ref{table2}:
(a) for two added holes and (b) for two removed holes ($=$ two added electrons) with respect to the undoped ladder.}
\end{figure}

The charge gap~(\ref{eq:charge_gap}) is calculated by adding two holes or two electrons to the undoped ladders.
Thus to understand the nature of  low-energy charge excitations we can study
the corresponding variation of the local density 
\begin{equation}
\Delta \rho_i = \rho^{\prime}_i - \rho_i
\end{equation}
where $\rho_i$ is the density of the undoped ladder while $\rho^{\prime}_i $
designates the density of the ladder with two holes or two electrons added.
In Fig.~\ref{fig03} we show the normalized distribution
\begin{eqnarray}\label{eq:local_density}
\bar{\rho}_i = \frac{\Delta \rho_i}{\sum_j \vert  \Delta \rho_j\vert}
\end{eqnarray}
for a r-ladder as an example.
If the density variation has the same sign for all orbitals $i$, $\bar{\rho}_i = \Delta \rho_i/2$.
Clearly, added holes  primarily occupy $p$ orbitals while holes are mostly removed
from $d$ orbitals when electrons are added. Similar results are found for the t-ladders and g-ladders
as well as for the common ladders~\cite{Polat2025}.
Therefore, the charge gap, the ground-state charge distribution ratio, and the density variations of low-energy excitations
demonstrate that the (t,g,r)-ladders are charge-transfer insulators for parameters around the values given in Table~\ref{table2}.

\begin{figure}[t]
\includegraphics[width=0.48\textwidth]{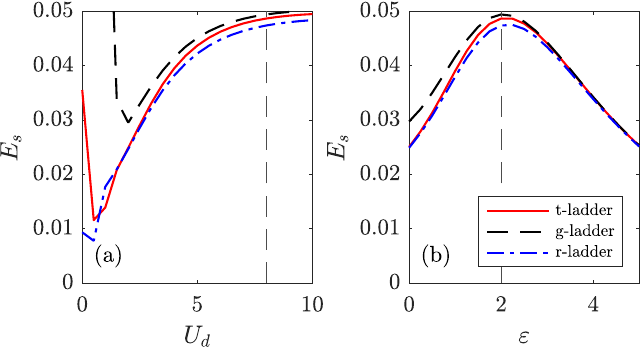}
\caption{\label{fig04} Spin gaps $E_s$ (\ref{eq:spin_gap}) as a function of the Hamiltonian parameters  $U_d$ (left panel) and $\varepsilon$ (right panel) for undoped (t, g, r)-ladders with a length of $L=24$.
The other Hamiltonian parameters are given in Table~\ref{table2}. Vertical dashed lines indicate the selected values  $U_d=8$ and $\varepsilon=2$, respectively.
}
\end{figure}

The spin gap can also be calculated from ground-state energies. It is given by
\begin{align}
E_s=E_0(N_\uparrow+1,N_\downarrow-1)-E_0(N_\uparrow,N_\downarrow) . \label{eq:spin_gap}
\end{align}
Figure~\ref{fig04} shows the spin gap of ladders with length $L=24$ as a function of the Hamiltonian parameters $U_d$ and $\varepsilon$.
The spin gaps are clearly much smaller than the charge gaps but these values are comparable
to those found in the common ladders~\cite{Jeckelmann1998c,Nishimoto2002b,Song2023,Song2021}.
However, the spin gaps 
are barely larger than the finite-size effects, which can be estimated from the results for $U_d=0$.
As discussed above the single-particle spectra of the t-ladder and the r-ladder are gapless
and thus their spin gaps should vanish for $U_d=0$ in the thermodynamic limit. 
By contrast, the noninteracting g-ladder has a significantly larger spin gap for small $U_d$  because of the band gap in its single-particle spectrum.

 \begin{figure}[b]
\includegraphics[width=0.48\textwidth]{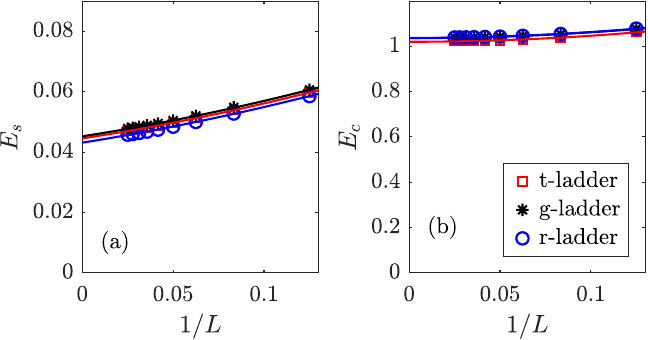}
\caption{\label{fig05} Convergence analysis of (a) spin gaps $E_s$ and (b) charge gaps $E_c$ as a function of the inverse of the ladder length $L$ in undoped  (t,g,r)-ladders  with  the parameter sets
from Table~\ref{table2}. 
Lines represent quadratic fits of the data for $40 \geq L \geq 16$.
}
\end{figure}

Therefore, we have analyzed the convergence of the spin gaps as a function of the ladder length using DMRG results up to $L=40$.
Some scalings are shown in Fig.~\ref{fig05}(a) for the (t, g, r)-ladders with their parameter sets in Table~\ref{table2}. Clearly, the 
spin gaps remain finite in the thermodynamic limit. 
Moreover, they are compatible with experimental values for cuprate ladder materials~\cite{Popovi2000,Gozar2001}
assuming $t_{dpx} \approx$ 1eV.
Figure~\ref{fig05}(b) confirms that finite-size effects are negligible for the
corresponding charge gaps in undoped ladders.

\begin{figure}
\includegraphics[width=0.48\textwidth]{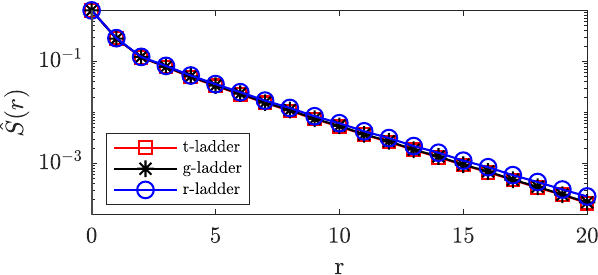}
\caption{\label{fig06} Spin correlation functions for holes in $d$-orbitals of undoped (t,g,r)-ladders with length $L=40$ and the parameter sets from Table~\ref{table2}. 
Lines are guides for the eye.}
\end{figure}

Finally, we have verified that the undoped (t,g,r)-ladders have short-range antiferromagnetic order between holes on $d$ orbitals.
For this purpose we have calculated the spin-spin correlation function
\begin{align}
S(r)= \langle S^{z}_{i}S^{z}_{j} \rangle
\end{align}
where $S^{z}_{i} = n^d_{i,\uparrow}-n^d_{i,\downarrow}$, $i$ is the index of a $d$ orbital in the middle of the ladder, $j$ designates
a $d$ orbital on the same leg, and $r$ is the distance  between these two orbitals (in unit of the unit cell of the CuO$_2$  lattice).
In Fig.~\ref{fig06}, we show the normalized and staggered correlation functions
\begin{align}
\hat{S}(r)=\frac{S(r)}{S(r=0)}\cdot (-1)^r
\end{align}
for the (t,g,r)-ladders.
As expected for spin-gapped two-leg ladders, $\hat{S}(r)$ is positive and decays exponentially with the distance $r$. 

In summary, the three ladder structures in Fig.~\ref{fig01}(b) to (d) are charge-transfer insulators with short-range antiferromagnetic order 
for the parameter sets in Table~\ref{table2}. Thus the common ladder structures and the novel ones have similar properties 
when undoped. Although the Hamiltonian parameters have been  adjusted, they all lie in a range that is
reasonable for superconducting cuprates.

\section{Doped ladders \label{sec:doped}}

 \begin{figure}[b]
\includegraphics[width=0.48\textwidth]{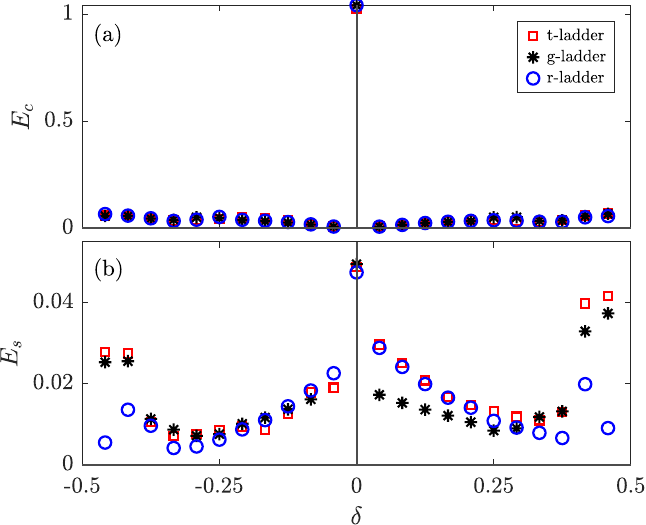}
\caption{\label{fig07} (a) Charge gap $E_c$ and (b) spin gap $E_s$ as functions of doping $\delta$ in (t,g,r)-ladders  with a length $L=24$ and the parameter sets from Table~\ref{table2}.
}
\end{figure}

The Emery model on the common ladder structures exhibits a Luther-Emery phase with PBE and enhanced pairing correlations
at low doping in the parameter range that is relevant for the superconducting cuprates~\cite{Jeckelmann1998c,Nishimoto2002b,Song2023,Song2021,Polat2025}.
However, the rung reflection symmetry and the Fermi points play an important role for the occurrence of the
Luther-Emery phase in two-leg ladders~\cite{Giamarchi2003}. Thus it is not clear if this phase can occur
in ladders that  breaks the rung reflection symmetry or have a reduced first Brillouin zone.
In this section we show that the doped  (t,g,r)-ladders can nevertheless exhibit a Luther-Emery phase.

 \begin{figure}
\includegraphics[width=0.48\textwidth]{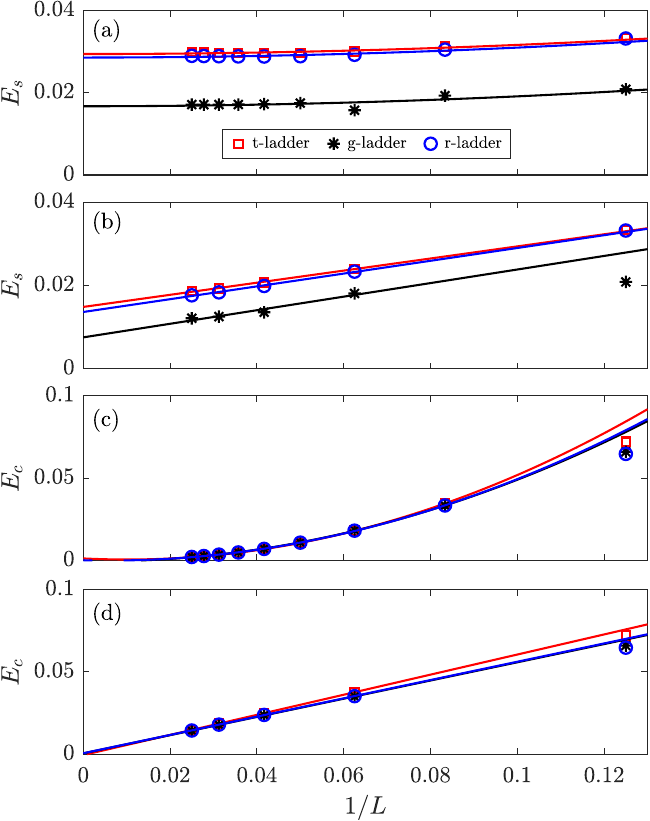}
\caption{\label{fig08} Convergence analysis of spin and charge gaps  as a function of the inverse of the ladder length $L$ 
in hole-doped (t,g,r)-ladders with the parameter sets in Table~\ref{table2}.
(a) Spin gaps $E_s$ for two doped holes  $(\delta=1/L)$, (b) spin gaps for a fixed doping $\delta=1/8$, (c) charge gaps $E_c$ for two doped holes  $(\delta=1/L)$, and (d) charge gaps for a fixed doping $\delta=1/8$.
Lines represent polynomial fits of the data  for $40 \geq L \geq 16$.
 }
\end{figure}

The evolution of the charge and spin gaps upon doping is shown in Fig.~\ref{fig07} for the range of $-0.5<\delta<0.5$.
The charge gap drops from values around $t_{dpx} = 1$ at $\delta=0$ to much smaller values for $\delta \neq 0$.
On the contrary, the spin gaps remain of the same order of magnitude for all dopings shown.
This suggests that the doped ladders have gapless charge excitations but only gapped spin excitations
as expected for a Luther-Emery phase.

However, the charge and spin gaps shown in Fig.~\ref{fig07} have been calculated in finite ladders of length $L=24$
and they are again of the order of the finite-size effects for $\delta \neq 0$.
As we can only calculate the gaps of  ladders up to a length  $L=40$ with our DMRG code,
we cannot analyze the finite-size scaling at fixed low doping. Thus we have analyzed 
finite-size scalings for two doped holes or electrons corresponding to an infinitesimal doping $\vert \delta \vert = 1/L \rightarrow 0^+$ in the thermodynamic limit $L \rightarrow \infty$
as well as for a fixed but relatively high doping $\vert \delta \vert = 1/8$. 
Some results are shown in Fig.~\ref{fig08}
for each ladder structure with their parameter sets from Table~\ref{table2}.
Clearly, the spin gaps remain finite in the thermodynamic limit in Fig.~\ref{fig08}(a) and (b)
while the charge gaps extrapolate to 0 for $1/L \rightarrow 0$ in Fig.~\ref{fig08}(c) and (d).
Therefore, our results indicate a phase, where all spin excitation modes are gapped but (at least) one charge excitation mode is gapless
in the thermodynamic limit,  which is compatible with a Luther-Emery phase.

\begin{figure}
\includegraphics[width=0.48\textwidth]{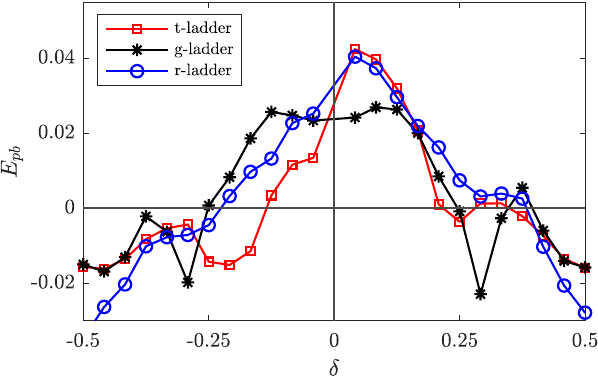}
\caption{\label{fig09} Pair binding energy $E_{pb}$ as functions of doping $\delta$ in doped (t,g,r)-ladders with length $L=24$ and their respective parameter sets in Table~\ref{table2}.
}
\end{figure}

\begin{figure}
\includegraphics[width=0.48\textwidth]{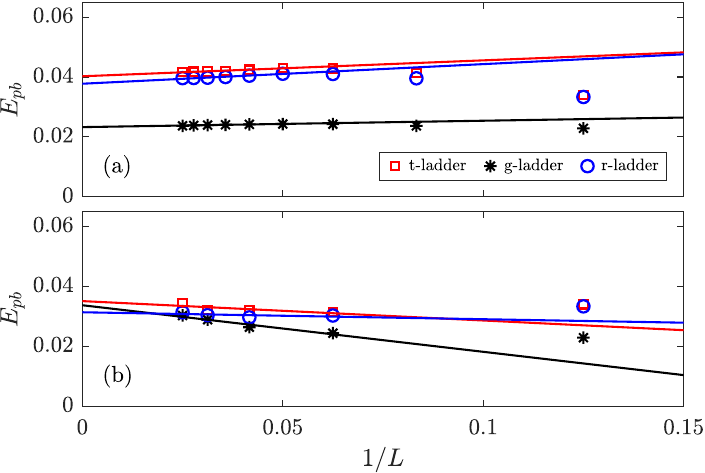}
\caption{\label{fig10} Convergence analysis of the pair binding energy $E_{pb}$ as a function of the inverse of the ladder length $L$
in hole doped (t,g,r)-ladders  with the parameter sets from Table~\ref{table2}. (a) For 
two doped holes $(\delta=1/L)$ and (b) for a fixed doping $\delta=1/8$. 
Lines represent linear fits of the data  for $40 \geq L \geq 16$.  
}
\end{figure}

The PBE~(\ref{eq:Epb}) is our primary tool for ascertaining the occurrence of pairing in ladder systems.
Figure~\ref{fig09} shows the PBE in the (t,g,r)-ladders for dopings in the range $-0.5<\delta<0.5$. 
In most cases the PBE values are positive at low absolute doping $\vert \delta \vert$ but vanish for larger $\vert \delta \vert$.
Moreover, they are similar in size to the spin gaps, as found for common ladders in previous works~\cite{Jeckelmann1998c,Nishimoto2002b}.
Consequently, these PBE are of the order of the finite-size effects discussed above for the spin gaps.
Thus we have also investigated their scaling as a function of the ladder length for two doped particles
and at fixed doping $\vert \delta \vert=1/8$. Some scalings are shown in Fig.~\ref{fig10} for the (t,g,r)-ladders with 
their selected parameters from Table~\ref{table2}. Obviously, the PBE converge to small but finite values for $1/L \rightarrow 0$.
The standard deviations for the PBE in the thermodynamic limit are significantly smaller than
the extrapolated values. For instance, they 
are smaller than $2 \cdot 10^{-3}t_{dpx}$ for the data presented in Fig.~\ref{fig10}.
As we observed for the common ladders~\cite{Polat2025}, the convergence is fast and regular because oxygen sites are included
at both ends of all ladder structures, see Fig.~\ref{fig01}.

Note that the calculated PBE and spin gaps may seem to be small compared
to the bare Hamiltonian parameters $U_d, \varepsilon,$ and $t_{dp}$. However, the effective band width
for spin and charge excitations is only of the order of $W \sim 0.5t_{dpx}$ in Emery ladders~\cite{Jeckelmann1998c}.
Thus the ratio between gaps ($\Delta$) and band width  ($W$) is not small 
and consequently finite-size effects decay on a scale set by $\xi \sim W/\Delta$, which is sufficiently smaller than
the system sizes $L$. For instance, the spin correlations in Fig.~\ref{fig06} decay exponentially with a correlation
length $\xi$ between 2 and 3 (in unit of the distance between two copper sites), which confirms the small ratio between band widths and gaps 
in Emery ladders.

The asymmetry between the PBE for electron and hole doping seen in Fig.~\ref{fig09} is a consequence of our approach.
We have searched for Hamiltonian parameters that yield significant positive PBE for hole-doped ladders
because we are mostly interested in hole-doped superconducting cuprates.
This choice often but not always leads to positive PBE for electron-doped ladders too, as shown by the results in Fig.~\ref{fig09}.
An obvious remedy would be to select other Hamiltonian parameters based on the PBE for two doped electrons.

Overall the size of the PBE and its evolution in the (t,g,r)-ladders is comparable to the observations made for the common ladders~\cite{Jeckelmann1998c,Polat2025}.
To support the evidence for pairing provided by the positive PBE, we have computed pairing correlation functions
as done in earlier works~\cite{Nishimoto2002b,Song2021,Song2023,Jiang2023a,Jiang2023b,Yang2024,Polat2025}.
Here we have focused on the correlations between rung singlet pairs made either of two $d$ orbitals or of one $d$ and one $p_y$ orbital
because we found previously in the common ladder structures~\cite{Polat2025} that  they were the slowest decaying correlations for electron and hole doping, respectively.
Their generic form is
\begin{equation}\label{eq:correlation}
P(r) = \frac{1}{2} \left(  \left\langle  \Delta_x^{\dagger}  \Delta^{\phantom{\dagger}}_{x^{\prime}} \right\rangle + \left\langle  \Delta^{\phantom{\dagger}}_x \Delta_{x^{\prime}}^{\dagger}  \right\rangle  \right) 
\end{equation}
where either
\begin{equation}\label{eq:correlation_delta}
 \Delta_x^{\dagger} = \frac{1}{2} \left(  d^{\dagger}_{j\uparrow} d^{\dagger}_{j'\downarrow} - d^{\dagger}_{j\downarrow} d^{\dagger}_{j'\uparrow} \right) 
\end{equation}
creates a singlet hole pair
in two $d$ orbitals $j=(x,1)$ and $j^{\prime}=(x,2)$ of the same rung $x$,
or 
\begin{equation}\label{eq:correlation_delta2}
 \Delta_x^{\dagger} = \frac{1}{2} \left(  d^{\dagger}_{j\uparrow} p^{\dagger}_{yj'\downarrow} - d^{\dagger}_{j\downarrow} p^{\dagger}_{yj'\uparrow} \right) 
\end{equation}
creates a singlet hole pair in a $d$ orbital $j=(x,1)$  and the middle $p_y$ orbital of the same rung $x$.
In both cases $r$ is the distance between the rungs $x$ and $x^\prime$. 
The expectation value in~(\ref{eq:correlation}) is calculated for the doped ground state of the Hamiltonian~(\ref{eq:hamiltonian}).

 \begin{figure}
\includegraphics[width=0.46\textwidth]{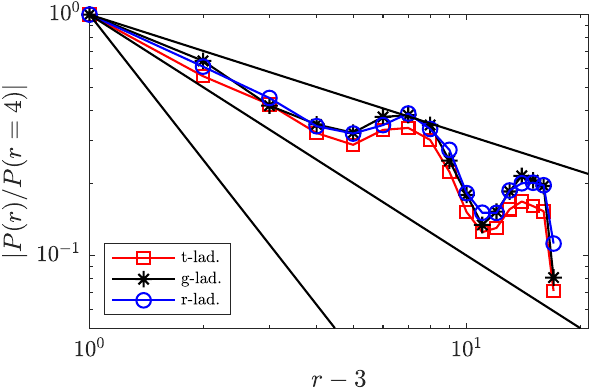}
\caption{\label{fig11} Pairing correlation functions~(\ref{eq:correlation})  for singlet pairs on $d$-$p_y$ rungs (\ref{eq:correlation_delta2}) in hole-doped (t,g,r)-ladders with length $L=40$, doping $\delta=1/8$,  and the parameter sets in Table~\ref{table2}. 
Three black lines show a power-law decay $(r-3)^{-\alpha}$  for $\alpha=1/2, 1$, and $2$.
}
\end{figure}

We have found that pairing correlations $P(r)$ are usually enhanced in the (t,g,r)-ladders when their PBE is positive.
Clear results are shown in Fig.~\ref{fig11} for the $d$-$p_y$ rung correlations~(\ref{eq:correlation_delta2}) in hole-doped ladders
with length $L=40$.
The correlation functions are normalized with their value at distance $r=4$ and shown on a double logarithmic scale
as a function of $r-3$ to avoid large short-range oscillations that would skew the analysis of the asymptotic behavior for $r \gg 1$.
The correlation functions seem to decay as $r^{-1}$ or slower and thus are clearly enhanced in comparison to noninteracting ladders,
where $P(r) \sim r^{-2}$.

The $d$-$d$ rung correlations~(\ref{eq:correlation_delta}) are enhanced in the common ladders upon electron doping
when the PBE is positive~\cite{Nishimoto2002b,Polat2025}. 
The picture is less clear for the (t,g,r)-ladder structures discussed here.
We have found that the r- and g-ladders usually exhibit enhanced correlations between $d$-$d$ rung pairs with a $r^{-1}$ decay.
These correlations seem to decay faster in the t-ladder but we have renounced drawing a conclusion 
about a quasi-long-range pairing order due to the limited system size $L \leq 40$ that we can calculate with DMRG.
As noted in Sec.~\ref{sec:model}, 
the t-ladder is related to the asymmetric Hubbard ladder, in which the observed pairing could not be explained by the formation of rung pairs alone~\cite{Abdelwahab2023}
due to the absence of the reflection symmetry.
Thus it is possible that a quasi-long-range pairing order only becomes clearly visible in electron-doped t-ladders with more complex pair structures than~(\ref{eq:correlation_delta}) and~(\ref{eq:correlation_delta2})
due to their lower symmetry.

\begin{figure}
\includegraphics[width=0.44\textwidth]{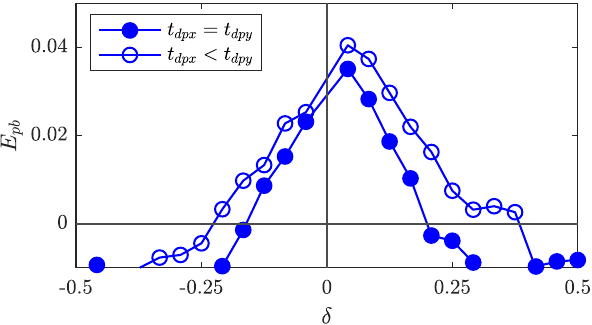}
\caption{\label{fig12} Pair binding energy $E_{pb}$ as function of doping $\delta$ in r-ladders with length $L=24$.
The empty blue circles are the results for the parameter set in Table~\ref{table2}, which are already shown in Fig.~\ref{fig09}.
Solid blue circles show the pair binding energy for the same parameters $\varepsilon$ and $U_d$ but $t_{dpx}=t_{dpy}$.
}
\end{figure}

\begin{figure}
\includegraphics[width=0.44\textwidth]{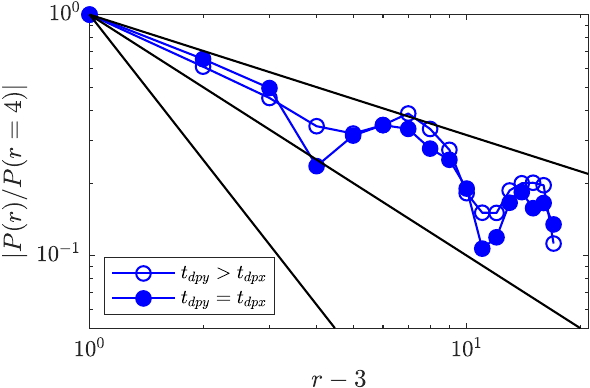}
\caption{\label{fig13} Pairing correlation functions  (\ref{eq:correlation})  for singlet pairs on $d$-$p_y$ rungs (\ref{eq:correlation_delta2}) 
in hole-doped  r-ladders with length $L=40$ and doping $\delta=1/8$.
The empty blue circles are the results for the parameter set in Table~\ref{table2}, which are already shown in Fig.~\ref{fig11}.
Solid blue circles show the pairing correlations for the same parameters $\varepsilon$ and $U_d$ but $t_{dpx}=t_{dpy}$.
Three black lines show a power-law decay $(r-3)^{-\alpha}$  for $\alpha=1/2, 1$, and $2$. 
}
\end{figure}

In all Emery ladders that we have investigated, the PBE is not maximal at $t_{dpy}=t_{dpx}$. The four-chain tube is a special case
because the periodic boundary condition in the rung direction strongly affects the noninteracting band structure and
pairing occurs only for $t_{dpy} < t_{dpx}$ \cite{Polat2025}. In the other five ladder structures, allowing $t_{dpy}>t_{dpx}$ leads to larger PBE.
However, superconducting cuprates correspond to a 2D Emery model that is invariant under 90 degree rotations
and thus to $t_{dpy}=t_{dpx}$. If we want the ladder structures in Fig.~\ref{fig01}(b) to (d) to be supercells of the 2D CuO$_2$ lattice,
$t_{dpy}=t_{dpx}$ must be used.
Consequently, we have also checked that positive PBE and enhanced pairing correlations occurs in (t,g,r)-ladders with $t_{dpy}=t_{dpx}$.
As an example, Fig.~\ref{fig12} compares the doping dependence of the PBE in  r-ladders with $t_{dpy} > t_{dpx}$ and with $t_{dpy}=t_{dpx}$
using the same parameters $\varepsilon$ and $U_d$ from Table~\ref{table2}.
Clearly, there is a region with positive PBE at low doping $\vert\delta \vert$  in the r-ladder with $t_{dpy}=t_{dpx}$ although this region is narrower than for a r-ladder with $t_{dpy}>t_{dpx}$.
Moreover, we see in Fig.~\ref{fig13} that the decay of pairing correlation functions is similar for both sets of parameters.

 In summary, the (t,g,r)-ladder structures exhibit a Luther-Emery phase with a vanishing charge gap, a positive PBE of the same order as the spin gap, and
 enhanced pairing correlations. This phase occurs at low doping for a range of parameters that is physically reasonable for superconducting cuprates.
 The pairing strength and its evolution with doping are similar for the novel and the common ladder structures.
 For a fixed set of Hamiltonian parameters the pairing strength varies with the ladder structure showing the influence of the oxygen sites.
 Consequently, one has to select different parameters for each ladder structure to obtain similar values of gaps or PBE.
 Pairing is also found for $t_{dpy}=t_{dpx}$, although it is weaker than for $t_{dpy} > t_{dpx}$,
 so that the Luther-Emery phase occurs when the (t,g,r)-ladders can be considered as supercells of the 2D CuO$_2$ lattice.

\section{Charge distribution  \label{sec:charge}}

The question of the charge distribution between Cu and O atoms in the superconducting cuprate layers has attracted a lot of attention recently.
Experimental data reveal a correlation between the average occupation $n_p$ of the planar $p$ orbitals of O atoms and the critical temperature 
for superconductivity when varying the chemical doping or the applied pressure~\cite{Rybicki2016,Jurkutat2023}.
Several theoretical studies have been carried out to determine the charge distribution between Cu and O orbitals and verify the observed correlation
between $n_p$ and  the strength of the superconducting state. They include
studies of the electronic version of the Emery model on a CuO$_2$ lattice using cluster methods~\cite{Kowalski2021,St-Cyr2025,Reaney2025}
and a study of the hole Hamiltonian~(\ref{eq:hamiltonian}) on lattices with up to $12\times 12$ sites using a quantum Monte Carlo method~\cite{Peng2025}.

\begin{figure}
\includegraphics[width=0.48\textwidth]{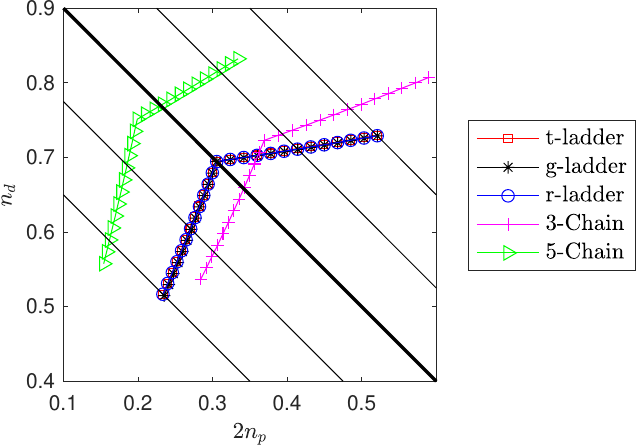}
\caption{
\label{fig14}
Relation between the hole occupation $n_d$ of copper $d$ orbitals and the hole occupation $2n_p$ of oxygen $p$ orbitals on a ladder with length 
$L=24$ when the doping $\delta$ is varied. Results are shown for the (t,g,r)-ladder structures and two common ladder structures (3-chain and 5-chain) using the  parameter sets from Table \ref{table2}.
For (t,g,r)-ladders the thick diagonal line corresponds to undoped ladders ($\delta=0$) while the lower two diagonal lines correspond to constant electron dopings 
($\delta = -1/8$ and $-1/4$) and the upper two diagonal lines indicate constant hole dopings ($\delta = +1/8$ and $+1/4$).
}
\end{figure}

An important issue is the relation between  $n_p$ and the average occupation $n_d$ of the planar $d$ orbitals of Cu atoms. In the ladder models
considered here
they fulfill the sum rule
\begin{equation}
n_d + \gamma n_p = 1+ \delta
\end{equation}
with the  ratio $\gamma$ of $p$ to $d$ orbitals (see Sec.~\ref{sec:undoped})
and the hole doping rate $\delta$.
In the superconducting cuprates $\gamma=2$ and thus recent studies often discuss $n_d$ as a function of $2n_p$  ~\cite{Jurkutat2023,St-Cyr2025,Reaney2025,Peng2025}.
Figure~\ref{fig14} shows the relation between $n_d$ and $2n_p$ in the (t,g,r)-ladder structures with the Hamiltonian parameters from Table~\ref{table2} when the doping rate $\delta$ is varied.
We clearly see that for electron doping most holes are removed from the $d$ orbitals while for hole doping most holes occupy $p$ orbitals, resulting in
two lines with different slopes for $\delta < 0$ and $\delta > 0$.
This is the behavior expected for a charge-transfer insulator. In the extreme case $t_{dpx}=t_{dpy}=0$ with $U_d > \varepsilon$,
one would obtain a vertical line for $\delta < 0$ and a horizontal one for $\delta > 0$.
By contrast, a pure Mott-Hubbard insulator yields a straight line (around $\delta=0$), which becomes vertical in the extreme case $t_{dpx}=t_{dpy}=0$ for $U_d < \varepsilon$.
Note that this behavior does not give any evidence about the existence of a Luther-Emery phase. We have indeed verified that the same behavior occurs in the Luttinger liquid phase that 
is reached when we increase the ratio $t_{dpy}/t_{dpx}$~\cite{Polat2025}.

For comparison,  in Fig.~\ref{fig14} we also show the results obtained for the three-chain and five-chain ladders with their respective parameter sets in Table~\ref{table2}.
First, we note that all results are qualitatively similar with two linear segments, which again just reflects charge-transfer insulators.
However, there are clear quantitative differences between the (t,g,r)-ladder structures and the two common ones.
In particular, the corners between the two linear segments do not lie on the main diagonal for the three and five-chain ladders although they correspond to undoped ladders.
This discrepancy can be easily corrected if we plot $n_d$ as a function of $\gamma n_p$ with $\gamma = 3/2$ and $5/2$, respectively.
However, this rescaling also changes the slope of the linear segments and thus impede the analysis of the relations between charge distribution, PBE, and Hamiltonian parameters. 
Therefore, preserving the ratio of Cu to O atoms is a clear advantage of  the (t,g,r)-ladder structures over the common ones when studying these relations.

\begin{figure}
\includegraphics[width=0.48\textwidth]{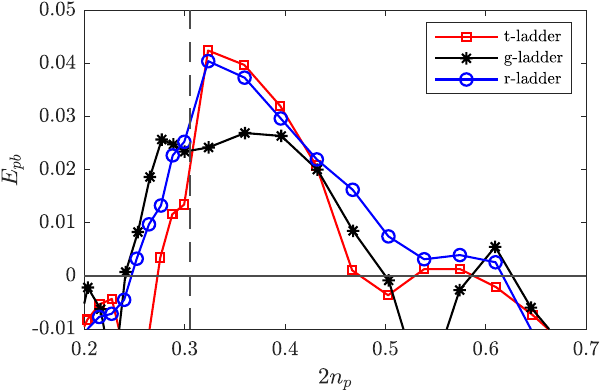}
\caption{\label{fig15} Relation between the PBE $E_{pb}$ and the oxygen occupation $2n_p$ when the doping $\delta$ is varied in (t,g,r)-ladders with $L=24$ and the parameters from Table~\ref{table2}.
The vertical dashed line shows $2n_p$ for undoped ladders ($\delta =0$) with electron doping ($\delta <0$) on the left-hand side and hole doping on the right-hand side ($\delta > 0$).
}
\end{figure}

To illustrate the pairing strength in relation to the charge distribution, Fig.~\ref{fig15} presents  a parametric plot of the PBE and the oxygen occupation $2n_p$ when the doping is varied. 
The same data are shown as $E_{pb}$ versus $\delta$ in Fig.~\ref{fig09}.  As explained in Sec.~\ref{sec:hamiltonian} the undoped ladders correspond to  $2n_p \approx 0.3$ due to our choice of 
Hamiltonian parameters (Table~\ref{table2}),
in agreement with various superconducting cuprates such as YBa$_2$Cu$_3$O$_{6+\delta}$.  In Fig.~\ref{fig15} we observe a maximum of the PBE as a function of the oxygen occupation in the range $0.3 \alt 2n_p \alt 0.4$.
This behavior is similar to the parabolic-like dependence of the critical temperature on the oxygen occupation
that is found experimentally in superconducting cuprates~\cite{Rybicki2016,Jurkutat2023}.

\begin{figure}
\includegraphics[width=0.37\textwidth]{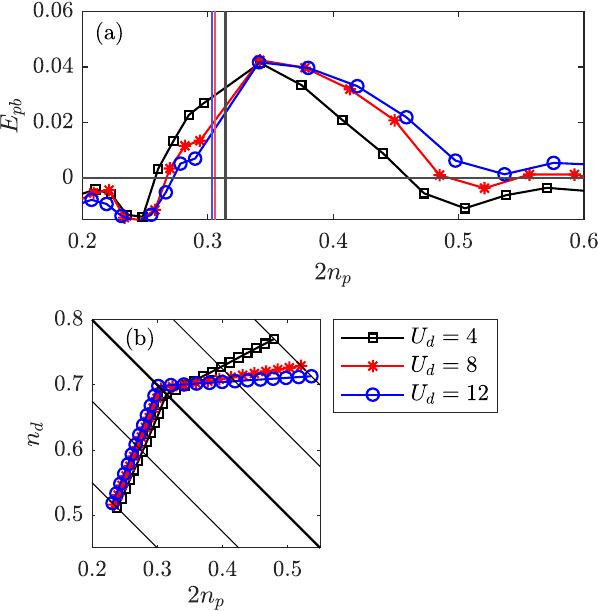}
\caption{\label{fig16} 
(a) Relation between the PBE $E_{pb}$ and the oxygen occupation $2n_p$ when doping is varied as in Fig.~\ref{fig15} but for a t-ladder with various interaction strengths $U_d$.
 The vertical lines indicate $2n_p$ for undoped ladders. (b) Relation between the copper occupation $n_d$ and the oxygen occupation $2n_p$ as in Fig.~\ref{fig14} 
 but for the same t-ladder as in (a).
}
\end{figure}

Previous studies have shown that the relation between Hamiltonian parameters and pairing strength is quite complex in
Emery ladders~\cite{Jeckelmann1998c,Nishimoto2002b,Nishimoto2009,Song2021,Song2023,Jiang2023a,Jiang2023b,Yang2024,Polat2025}.
In contrast, the charge distribution between Cu and O orbitals seems to vary rather systematically.
As a first example, Fig.~\ref{fig16}(a) presents a parametric plot of the PBE and the oxygen occupation when the doping $\delta$ varies for three values of the interaction $U_d$ in a t-ladder. 
We observe that a larger  $U_d$ increases the pairing strength in hole doped ladders but reduces it in electron doped ladders. However,
the overall relation between $E_{pb}$ and $2n_p$ is qualitatively similar for the range $4 \leq U_d \leq 12$.
In particular, the oxygen occupation  of the undoped ladder varies little with $U_d$ in that range, as shown by the vertical lines in Fig.~\ref{fig16}(a).
Figure~\ref{fig16}(b) shows the evolution of the charge distribution between $p$ and $d$ orbitals as $\delta$ varies for the same ladders.
We see that the slope of $n_d$ versus $2n_p$ is slightly lower for stronger $U_d$ in hole-doped ladders but little change occurs for electron doping.
Clearly,  doped holes tend to occupy $p$ orbitals to avoid the energy cost $U_d$ of  doubly-occupied $d$ orbitals and this trend
becomes stronger for higher $U_d$. In contrast, electron doping corresponds to removing holes from singly-occupied $d$ orbitals and thus the relation between 
$n_d$ and $2n_p$ remains largely
unaffected by the value of $U_d$.

\begin{figure}
\includegraphics[width=0.37\textwidth]{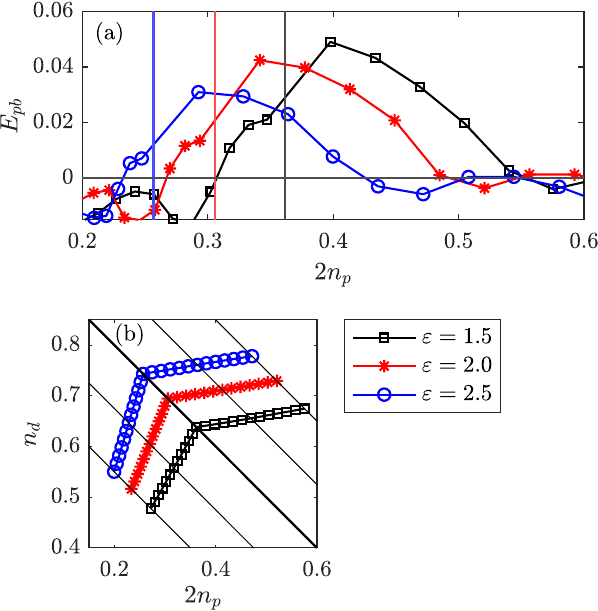}
\caption{\label{fig17} 
(a) Parametric plot of the PBE $E_{pb}$ and the oxygen occupation $2n_p$ when doping $\delta$ is varied as in Fig.~\ref{fig15} but for a t-ladder with various energy differences  $\varepsilon$
between oxygen $p$-orbitals and copper $d$-orbitals.
 The vertical lines indicate $2n_p$ for undoped ladders. (b) Copper occupation $n_d$ as a function of oxygen occupation $2n_p$ as in Fig.~\ref{fig14} 
 but for the same t-ladder as in (a).
}
\end{figure}

As a second example, Fig.~\ref{fig17}(a) presents another parametric plot of the PBE versus the oxygen occupation in a t-ladder
but
for three values of the energy difference $\varepsilon$  between  $p$ and $d$ orbitals.
We observe an increase of the maximal PBE when $\varepsilon$ becomes smaller but the corresponding
occupation $2n_p$ of the $p$ orbitals also shifts toward larger values. This shift also occurs in the undoped ladders
as shown by vertical lines in Fig.~\ref{fig17}(a).
Figure~\ref{fig17}(b) shows the evolution of the charge distribution between $p$ and $d$ orbitals for the same ladders.
As expected, we observe a shift  toward a higher occupation of $d$ orbitals when $\varepsilon$ is larger.  
The slope of $n_d$ as a function of $2n_p$ becomes smaller for hole doping but  larger for electron doping with increasing $\varepsilon$,
and thus tends toward the behavior predicted for a charge-transfer insulator in the limit $U_d > \varepsilon \gg t_{dpx},t_{dpy}$.

We observe a similar dependence of the charge distribution as a function of the Hamiltonian parameters in all three (t,g,r)-ladder structures. 
(Nevertheless, the pairing strength depends on the structure as illustrated by the PBE values in Table~\ref{table2}.)
A similar behavior of the charge distribution is observed in 2D Emery clusters~\cite{Peng2025}. 
Our results also agree qualitatively with the findings reported for the electronic version of the 2D Emery model~\cite{St-Cyr2025,Reaney2025}.
However, when comparing to these results, one has to take into account
that varying $U_d$ at fixed $\varepsilon$  in our hole Hamiltonian corresponds to varying $U_d$ at fixed $U_d-\Delta$ in the electronic Hamiltonian,
where $\Delta$ is the energy difference between one electron in a $p$ orbital  and one electron in a $d$ orbital.
These findings suggest that material-specific parameters for the Emery model could be determined by comparison 
 with first-principles simulations or experimental data for $n_d$ and $n_p$ \cite{Jurkutat2023}. Indeed, 
 the charge distribution of the undoped system determines the energy difference $\varepsilon$ between $p$ and $d$ orbitals
 while the slope of $n_d$ versus $2n_p$  upon hole doping determines the interaction $U_d$. 
 This comparison is only possible for Emery ladders preserving the Cu to O ratio, such as the ones shown in Fig.~\ref{fig01}(b)-(d).

\section{Conclusion \label{sec:conclusion}}

Studying strongly-correlated ladder systems is often seen as a way towards understanding their 2D  parent systems.
In the case of the Emery model for the CuO$_2$ plane, however,  the appropriate choice of ladder structures and model parameters
is an issue.
For simple models such as the single-orbital Hubbard model on a square lattice,  there is an unambiguous two-leg ladder sublattice
that  is a supercell of the 2D lattice.
For the Emery model, however,  we can define several different two-leg ladder structures
depending on the $p$ orbitals taken into account and the boundary conditions in the rung direction.
In our previous paper~\cite{Polat2025}  we showed that the ladder structures used in other works either did not have the right ratio of Cu to O atoms or did not
exhibit a Luther-Emery phase for isotropic hoppings.

In this work, we have investigated the Emery model on 
three ladder-like lattice structures with two legs of copper $d$ orbitals that
are supercells of the CuO$_2$ plane and thus preserve the ratio of copper to oxygen atoms.
For a range of parameters that is physically reasonable for cuprates these ladders describe the same physics 
as the common ladders studied in previous works despite their lower symmetry. They are charge-transfer insulators 
at a filling of one hole per Cu atom but exhibit a Luther-Emery phase with positive pair binding energy  and enhanced pairing correlations at low 
electron and hole doping. We have explicitly verified that pairing also occurs for isotropic hoppings. 
Overall, the previous ladder structures and the new ones exhibit qualitatively similar pairing properties.
The main differences are quantitative, e.g. the pair binding energies  are usually different for a given set of 
Hamiltonian parameters.

The main advantage of the (t,g,r)-ladder structures is that they can reproduce the distribution of charge between
the copper $d$ orbitals and the oxygen $p$ orbitals that is found in the Emery model on the 2D CuO$_2$ lattice.
Thus we can describe the relations between charge distribution, doping, pairing strength, and model parameters
in the Luther-Emery phase of doped charge-transfer insulators.
Moreover, comparison becomes possible with experimental results for the charge distribution.
This is not possible with the common ladder structures due to the incorrect Cu to O ratio
or the absence of the Luther-Emery phase for isotropic parameters.
More generally, the (t,g,r)-ladders  have the same (super) unit cell as the CuO$_2$ plane.
Thus they will allow us to study Emery ladders of varying width with DMRG to gain insight
into the properties of the isotropic 2D Emery model
as done for the one-band Hubbard model~\cite{LeBlanc2015}.
Therefore, we think that the ladder structures presented in this work constitute a better approach for studying 
pairing in the Emery model than the common ladder structures used so far if one aims to understand
the physics of high-temperature superconducting cuprates.

\acknowledgments

The calculations were carried out on the compute cluster at the Leibniz University of Hannover,
which is funded by the Deutsche Forschungsgemeinschaft (DFG, German Research Foundation),
project number  INST 187/742-1 FUGG.

\section*{Data availability statement}
 The data that support the findings of this article are openly available~\cite{dataset2026}.

%

\bibliographystyle{biblev4}
\bibliography{mybibliography}

\end{document}